\documentclass[9pt,twoside]{extarticle}

\usepackage{ucs}
\usepackage[utf8x]{inputenc}
\usepackage[english,francais]{babel}
\usepackage[T1]{fontenc}
\usepackage{graphicx}

\usepackage{hyperref}
\hypersetup{colorlinks=true, linkcolor=blue, filecolor=blue, pagecolor=blue, urlcolor=blue}
\usepackage{url}
\usepackage[authoryear]{natbib}

\usepackage[vmargin=28mm,hmargin=24mm]{geometry}\usepackage{cmbright}
\def\auteur{Martine \textsc{Hurault-Plantet}\up{1} -- Élie \textsc{Naulleau}\up{2} -- Bernard \textsc{Jacquemin}\up{3}}
\def\hbpauteur{Martine \textsc{Hurault-Plantet} \textit{et al.}}
\def\adresselabo{\up{1}LIMSI-CNRS UPR 3251, Orsay (France)\\\up{2}Semiosys, Les Sables d'Olonne (France)\\\up{3}CREM EA 3476, Metz, Mulhouse, Nancy (France)}

\def\courriel{\href{Martine.Hurault-Plantet@limsi.fr}{Martine.Hurault-Plantet\,@\,limsi.fr}\\\href{semiosys@semiophore.net}{semiosys\,@\,semiophore.net}\\\href{Bernard.Jacquemin@uha.fr}{Bernard.Jacquemin\,@\,uha.fr}}

\def\titre{GraphDuplex: visualisation simultanée de N réseaux couplés 2 par 2}

\def\mabiblio{JacqueminCoria2009}

\def\titrecourt{GraphDuplex}

\def\piedpage{\textit{CORIA 2009}, pp. 351--362}

\usepackage{fancyhdr}
\title{\titre}
\author{\auteur\\\adresselabo\\\courriel}
\date{}

\begin{document}

\maketitle

\begin{abstract}
L'analyse des réseaux sociaux fait un usage intensif d'outils de visualisation et, dans
le domaine de la recherche d'information, l'exploration visuelle de réseaux lexicaux est
utilisée comme une aide à la désambiguïsation ou au raffinement de la requête. Ces deux
types de réseaux se trouvent associés via Internet lorsqu'un contenu textuel est lié à une
activité sociale (méls, blogs, travail collaboratif). Dans cet article, nous présentons un
logiciel de visualisation simultanée de plusieurs réseaux, GraphDuplex, qui, combiné à des
méthodes statistiques, permet par exemple d'étudier conjointement un réseau social (ou
plusieurs) et son réseau lexical associé. GraphDuplex permet en particulier des requêtes
dynamiques inter-réseaux, entre les nœuds ou les liens des deux réseaux. \\
\textbf{Mots-clefs:} visualisation interactive, réseau social, réseau lexical, requête dynamique.
\end{abstract}

\selectlanguage{english}
\begin{abstract}
While social network analysis often focuses on graph structure of social actors, an
increasing number of communication networks now provide textual content within social
activity (email, instant messaging, blogging, collaboration networks). We present an open
source visualization software, GraphDuplex, which brings together social structure and
textual content, adding a semantic dimension to social analysis. GraphDuplex eventually
connects any number of social or semantic graphs together, and through dynamic queries
enables user interaction and exploration across multiple graphs of different nature. 
\end{abstract}
\selectlanguage{francais}

\section{Introduction}

L'exploration des propriétés d'un réseau se fait depuis longtemps à l'aide d'outils
de visualisation, en supplément ou en complément des outils mathématiques
classiques de la théorie des graphes. L'analyse des réseaux sociaux en particulier a
suscité de nombreuses recherches sur le sujet \citep{Freeman2004}. Les logiciels qui en
découlent proposent souvent, en plus de la visualisation proprement dite du réseau,
des méthodes permettant de mieux cibler ce qui est visualisé et d'avoir du réseau des
vues à la fois globales et locales. En particulier, \cite{BrandesWagner2004} présentent un
outil d'exploration d'un réseau social qui propose différents algorithmes de dessin de
graphe, privilégiant la facilité de lecture, adaptés aux réseaux de petite et moyenne
taille. Par ailleurs, \cite{PererShneiderman2006} ont développé un outil de visualisation
interactive qui intègre un ensemble de méthodes statistiques permettant de mettre en
valeur visuellement, par des couleurs ou des tailles de nœuds ou liens, des propriétés
particulières du réseau. L'intérêt s'est porté aussi sur la visualisation de très grands
réseaux \citep{BatageljMrvar2004}.

Moins présent sur ce thème, le domaine du traitement des données textuelles a
cependant intégré depuis longtemps des outils de visualisation pour la représentation
et l'analyse des réseaux lexicaux. \cite{Veronis2003} s'appuie sur la construction des
différentes composantes de forte densité d'un réseau lexical pour distinguer les
différents usages d'un même mot, dans le but d'utiliser son environnement lexical en
recherche d'information. La visualisation associée permet à l'utilisateur de naviguer
dans les thèmes liés au mot sélectionné. \cite{TunkelangAl1997} proposent une
méthode de raffinement d'une requête en recherche d'information par une navigation
dans le réseau lexical des documents guidée par les termes de la requête.

Cependant, on n'a accordé jusqu'à présent que peu d'attention à la visualisation
simultanée de plusieurs réseaux. Il s'agit de réseaux distincts dont les nœuds ont, en
plus de la relation qui les lie entre eux à l'intérieur de chaque réseau, une autre
relation qui les lie aux nœuds d'un autre réseau\footnote{Dans les graphes N-partis, on 
ne considère que les liens qui relient des nœuds appartenant à des ensembles différents. 
Ici, on considère aussi les liens entre nœuds d'un même ensemble.}. Les nœuds dans chaque réseau sont
de même nature, en revanche ils sont en général de natures différentes d'un réseau à
l'autre. Nous présentons dans ce qui suit le logiciel GraphDuplex\footnote{L’application et le code 
source Java sont téléchargeables sur \href{http://www.semiophore.net}{\url{http://www.semiophore.net}} avec une vidéodémonstration:
\href{http://semiosys.free.fr/video/graphduplex/}{\url{http://semiosys.free.fr/video/graphduplex/}}.} qui permet de
visualiser simultanément plusieurs réseaux qu'on peut coupler deux par deux. Ce
couplage permet des requêtes dynamiques. En effet, la sélection d'un nœud\footnote{On peut sélectionner 
plusieurs nœuds ou encore des liens.} de l'un
des réseaux couplés entraînera une modification visuelle de l'autre réseau, suivant la
relation qui les lie. Ce couplage est paramétrable et suivant le type des données
propose différentes relations (égalité, relations d'ordre, opérateurs ensemblistes\dots{}).
GraphDuplex visualise les réseaux dans des fenêtres séparées. Chaque fenêtre
possède un tableau de bord qui permet de régler différents paramètres de
visualisation. Les paramètres globaux sur le réseau sont l'algorithme de dessin du
graphe\footnote{Un ensemble de 10 algorithmes de dessin de graphe sont disponibles, 
apportés par les librairies Jung et Graphviz.}, la possibilité de déplacer chaque nœud du graphe ou le graphe dans son
ensemble, et enfin la possibilité de déplacer une loupe sur le graphe afin d'en
visualiser les détails. L'utilisation de cette loupe est particulièrement intéressante lors
de la visualisation de grands graphes. Les paramètres sur les nœuds (ou sur les liens)
comprennent l'affichage ou non des libellés, un choix de représentation des
propriétés du nœud (ou du lien) par des variations de couleur, de taille, ou encore par
des secteurs distributionnels. Les ajustements des paramètres de visualisation
permettent déjà de mettre en valeur certaines propriétés du réseau. Il s'y ajoute un
ensemble de possibilités de filtrage interactif qui permettent de ne visualiser que des
parties du réseau qui possèdent en commun une ou plusieurs propriétés données sur
les liens ou sur les nœuds. Par ailleurs les deux réseaux sont couplés, c'est-à-dire
qu'une action de clic sur l'un des éléments d'un réseau déclenche la mise en évidence
visuelle des éléments de l'autre réseau qui lui sont liés. Par exemple, cliquer sur un
nœud-individu du réseau social sélectionne visuellement le sous-réseau lexical du
vocabulaire de cet individu. Les données des réseaux sont chargées dans
GraphDuplex soit à partir de données sauvegardées dans GraphDuplex lors d'une
précédente session, soit, initialement, à partir des données d'une base de données, à
laquelle on accède par un fichier XML. Ce fichier contient les interrogations de la
base de données permettant de sélectionner les données qu'on veut visualiser.

Ce logiciel a été développé dans le cadre du projet 
Autograph\footnote{\href{http://autograph.fing.org/texts/PresentationAutograph}{\url{http://autograph.fing.org/texts/PresentationAutograph}}} sur la conception
d'outils de visualisation pour la gouvernance des communautés collaboratives sur
Internet, dont, en particulier, la communauté des contributeurs à Wikipédia. Les
productions écrites des wikipédiens ne se limitent pas aux articles encyclopédiques,
elles comprennent aussi toutes les discussions qui s'y réfèrent. D'une part cette
communauté constitue un réseau social, qui se subdivise en sous-communautés
suivant le type de lien social (par exemple, travail sur un même domaine, ou
participation à des tâches semblables d'administration de l'encyclopédie), et on peut
donc étudier ce réseau social en tant que tel. Et d'autre part cette communauté
partage un lexique, utilise, ou n'utilise pas, des termes semblables, et le réseau lexical
qui en découle peut également être étudié en tant que tel. Mais ces deux types de
réseaux sont également liés, chaque acteur du réseau social utilisant une partie du
lexique, et chaque mot du lexique étant utilisé par un ensemble d'acteurs. C'est
l'ensemble de ces propriétés, thèmes et sous-thèmes des différentes communautés,
qu'une interface de visualisation simultanée de plusieurs réseaux permet d'explorer.
Dans cet article, nous montrons les différentes possibilités du logiciel sur l'exemple
du réseau social des arbitres du Comité d'arbitrage de Wikipédia associé au réseau
lexical du vocabulaire qu'ils utilisent au cours des arbitrages.

\section{Le réseau des arbitres}

\subsection{L'arbitrage dans Wikipédia}

Wikipédia est un projet encyclopédique libre sur Internet \citep{ZlaticAl06},
couvrant tous les domaines du savoir, au sein de différentes communautés de langue
gérant leur projet de manière autonome. Ce savoir doit être présenté de manière
objective, suivant le principe de la neutralité de point de vue \citep{ViegasAl04}, et
l'ensemble du processus éditorial, de l'écriture des articles à l'organisation de la
macrostructure, est géré collectivement. Cela a impliqué la mise en place progressive
de divers instruments et procédures de régulation et de contrôle \citep{ViegasAl07}.
En particulier, un comité d'arbitrage a été mis en place pour régler les litiges d'édition
sévères entre contributeurs.

Dans l'instance française de Wikipédia, le comité d'arbitrage est un groupe
composé de sept membres de la communauté des contributeurs, élus par la
communauté pour une période de six mois. Ils sont chargés de recevoir les plaintes
des contributeurs en conflit ouvert (avec insultes dans les pages de discussion par
exemple), lorsque toutes les possibilités de médiation sont épuisées. Les
délibérations et les votes du comité d'arbitrage sont publics sur des pages de
Wikipédia qui leur sont dédiées\footnote{\href{http://fr.wikipedia.org/wiki/Wikipédia:Comité d'arbitrage/Arbitrage}{\url{http://fr.wikipedia.org/wiki/Wikipédia:Comité d'arbitrage/Arbitrage}}} 
et cherchent autant que possible l'unanimité,
privilégiant donc le consensus comme c'est la règle dans les articles. Les sanctions
votées par ce comité peuvent aller du blocage (interdiction technique et temporaire
de contribuer sur un ou plusieurs articles) au bannissement définitif (interdiction de
participer à tout contenu de Wikipédia).

\subsection{Réseau social et réseau lexical}

L'encyclopédie Wikipédia peut être librement téléchargée\footnote{Un fichier de sauvegarde 
de l'ensemble des données textuelles sous forme de données
MySQL compressées est mis à disposition sur \href{http://download.wikipedia.org/backup-index.html}{\url{http://download.wikipedia.org/backup-index.html}} 
et régulièrement mis à jour.} et exploitée. Nous
disposons de la sauvegarde de la base Wikipédia française réalisée le 2 avril 2006,
soit plus de 600 000 pages comprenant notamment près de 370 000 pages d'articles
auxquelles sont associées plus de 40 000 pages de discussion sur article.

Le corpus des arbitrages est constitué des quatre-vingts pages d'arbitrages de
notre base Wikipédia, suffisamment bien formées pour que l'information qu'elles
contiennent puisse être appréhendée automatiquement. Cent dix protagonistes et dix-
neuf arbitres ont confronté leurs avis au cours de ces débats. Chaque page d'arbitrage
doit respecter une structure donnée, qui consiste d'abord en une description du
conflit (les parties concernées, la nature du conflit), ensuite les preuves et arguments
des protagonistes, puis les commentaires des arbitres, et enfin le vote et la décision.

\begin{figure*}[!htb]
 \centering
 \includegraphics[width=.8\textwidth]{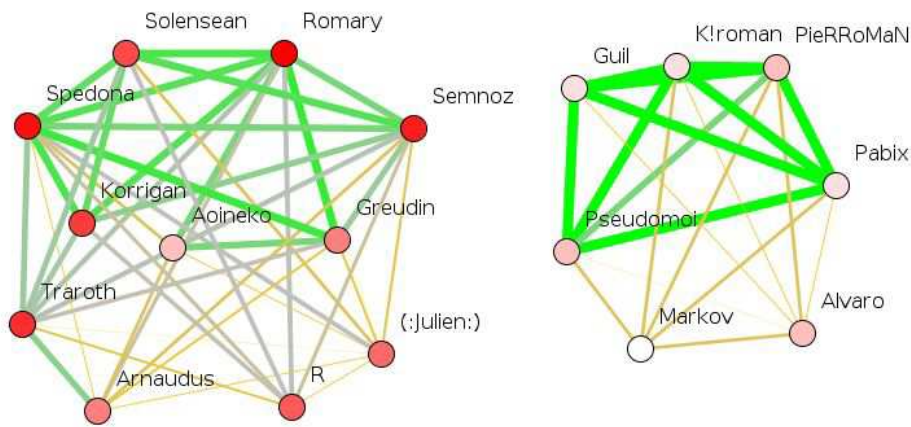}
 \caption{Le réseau des arbitres du Comité d'arbitrage de Wikipédia entre début 2001 et avril 2006.}
 \label{fig:1-arbitresNet}
\end{figure*}

Nous avons défini le lien social entre les arbitres par leur accord ou désaccord
dans les votes de décision d'arbitrage. La Figure \ref{fig:1-arbitresNet} montre le réseau des arbitres
visualisé par GraphDuplex. Les nœuds du réseau représentent les arbitres et les liens
entre les nœuds expriment leurs accords. Nous considérons qu'il y a un accord entre
deux arbitres lorsqu'ils votent tous deux de la même manière, c'est-à-dire pour, ou
bien contre, une proposition d'arbitrage.

Le poids sur chaque nœud correspond au nombre de participations de l'arbitre à
un vote. Sa valeur varie entre 2 et 87. Ces valeurs peuvent être visualisées par des
tailles différentes de nœud ou par des teintes différentes de couleur, comme sur la
Figure \ref{fig:1-arbitresNet} où la couleur du nœud est plus ou moins foncée suivant que l'arbitre a
participé à plus ou moins de votes. Dans chaque composante connexe, on remarque
un noyau d'arbitres ayant très souvent voté (nœuds plus foncés), et ayant été lors de
ces votes très souvent en accord les uns avec les autres (liens plus foncés et plus
épais). La plus grande des deux composantes correspond à un ensemble d'arbitres
qui ont arbitré pendant une grande partie de la période considérée (2001-2006).

Le poids sur le lien entre deux arbitres correspond à leur proportion d'accord sur
l'ensemble des votes auxquels ils ont participé ensemble. Sa valeur varie entre 25\%
et 100\%. Ces valeurs peuvent être visualisées par des tailles différentes de lien ou des
variations de couleur, ou les deux comme sur la Figure \ref{fig:1-arbitresNet} où la taille des liens varie et
leur couleur aussi varie en nuance, du foncé au clair, suivant la plus ou moins grande
proportion d'accord. L'accord est indépendant du nombre de participations aux
votes; Solensean par exemple, qui a participé à moins de votes que Traroth (nœud
plus clair), est en meilleur accord que lui avec les autres arbitres (liens plus foncés).

\begin{figure*}[!htb]
 \centering
 \includegraphics[width=.8\textwidth]{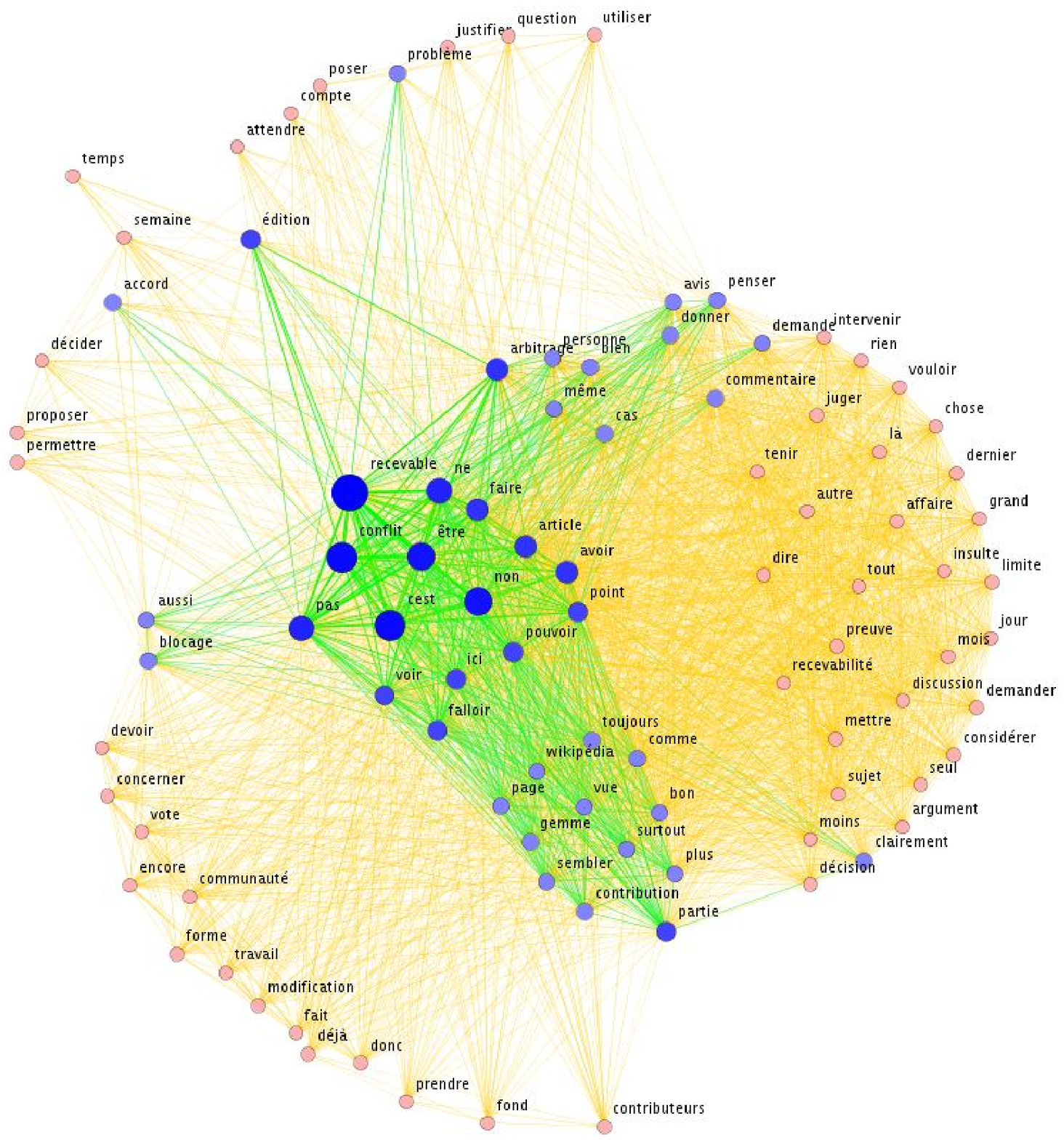}
 \caption{Le réseau lexical du vocabulaire des arbitres dans les arbitrages 
 Wikipédia entre début 2001 et avril 2006.}
 \label{fig:2-arbitresVoc}
\end{figure*}

Le réseau lexical associé est constitué de l'ensemble des noms, adjectifs, verbes,
et adverbes que les arbitres utilisent au cours de leurs débats dans les arbitrages.
Dans le réseau lexical, nous n'avons conservé que les termes dont la fréquence-
document dans ce corpus, c'est-à-dire le nombre d'arbitres qui utilisent ce terme, est
au moins égale à 10. Le réseau lexical résultant comporte 97 nœuds-termes\footnote{Le 
nombre de nœuds peut être de plusieurs milliers, et la fonction loupe permet alors de
visualiser les détails du réseau.} dont la
fréquence varie entre 10 et 18. Le poids de chaque nœud est la fréquence-document
du terme. Ce poids peut être visualisé par la taille et la couleur comme sur la
Figure \ref{fig:2-arbitresVoc} où la couleur est plus ou moins foncée suivant la plus ou moins grande
fréquence du nœud-terme.

Dans le réseau lexical, deux nœuds-termes sont reliés s'ils sont tous deux utilisés
par le même arbitre. Le poids du lien est d'autant plus fort que les deux termes sont
utilisés par un plus grand nombre d'arbitres, en valeur absolue (cooccurrence) ou
relativement à leur fréquence dans le corpus (mesure d'équivalence et information
mutuelle). La force du lien peut être visualisée par l'épaisseur et la nuance de
couleur, comme sur la Figure \ref{fig:2-arbitresVoc} où la couleur est plus ou moins foncée suivant le plus
ou moins grand nombre de cooccurrences entre les deux termes.

\section{La fenêtre de visualisation d'un réseau}

\subsection{Filtrage sur les nœuds et les liens du réseau}

Le logiciel GraphDuplex permet un ensemble de filtrages interactifs sur les
nœuds et sur les liens du réseau visualisé. Il est possible de filtrer les nœuds du
réseau par le nom du nœud, en sélectionnant ceux qu'on veut afficher ou masquer, ou
par le poids du nœud, en sélectionnant un seuil inférieur ou un seuil supérieur sur la
valeur du poids, ou les deux à la fois. On peut aussi filtrer les nœuds et les liens par
un autre attribut, dans notre cas les mots du lexique des arbitres.

\subsubsection{Filtrage sur le poids}

Le filtrage sur le poids des nœuds du réseau social permet de ne visualiser qu'une
certaine catégorie d'arbitres, liée au nombre de participations à des arbitrages.

\begin{figure*}[!htb]
 \centering
 \includegraphics[width=.8\textwidth]{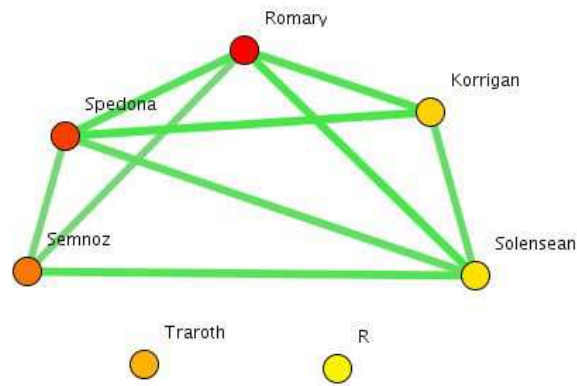}
 \caption{Le réseau des arbitres de Wikipédia filtré par leur nombre de 
 participations (au moins 50 votes au Comité d'arbitrage), et leur nombre d'accords 
 (au moins 75\% d'accord sur ces votes).}
 \label{fig:3-arbitresNetVotesAccords}
\end{figure*}

Nous nous sommes intéressés plus particulièrement aux arbitres les plus actifs.
Pour cela, nous filtrons le réseau sur le poids des nœuds, c'est-à-dire sur le nombre
de votes à des décisions d'arbitrage auxquels les arbitres ont participé. Par ailleurs, le
degré d'accord entre arbitres est encore plus visible lorsqu'on filtre les liens qui les
représentent par un seuil sur leur poids, c'est-à-dire sur la proportion d'accord entre
arbitres. La Figure \ref{fig:3-arbitresNetVotesAccords} montre le réseau obtenu en filtrant le réseau des arbitres sur une
participation à au moins 50 votes, et en ne conservant que les liens d'au moins 75\%
d'accord. Nous voyons donc que les arbitres Traroth et R ont un accord médiocre
avec les autres arbitres, alors que ces autres arbitres ont globalement un meilleur
accord entre eux. L'arbitre R est d'ailleurs resté peu de temps au Comité d'arbitrage.

\subsubsection{Filtrage sur le lexique}

\begin{figure*}[!htb]
 \centering
 \includegraphics[width=.8\textwidth]{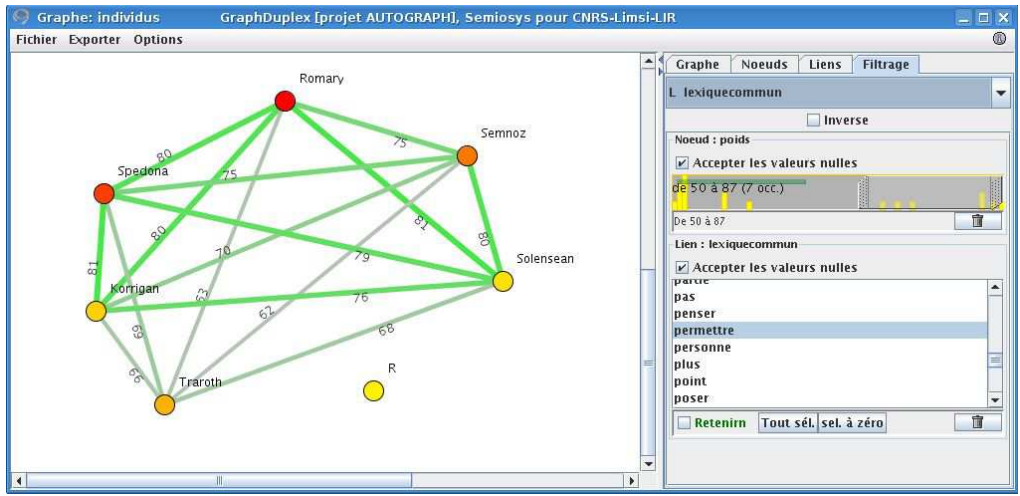}
 \caption{Le réseau des arbitres de Wikipédia ayant participé à au moins 50 votes 
 au Comité d'arbitrage, et ayant en commun le mot « permettre ».}
 \label{fig:4-arbitresNetPermettre}
\end{figure*}

Le logiciel Graphduplex nous permet aussi d'étudier le vocabulaire des arbitres,
et, en particulier, d'identifier quels termes les différencient. Un filtrage sur chacun
des termes du vocabulaire permet de visualiser les liens entre les arbitres qui
possèdent ce terme en commun dans leur vocabulaire\footnote{On peut faire le même filtrage 
de vocabulaire sur les nœuds-arbitres que sur les liens entre arbitres. Dans un cas on 
masque les arbitres qui n'utilisent pas un terme donné, dans l'autre on masque les liens 
entre arbitres qui n'ont pas ce terme en commun.}. Par exemple, nous voyons sur
la Figure \ref{fig:4-arbitresNetPermettre} que la sélection du mot \textit{permettre} supprime les liens reliant l'arbitre R aux
autres arbitres. Cela signifie que tous les arbitres, sauf R, utilisent le terme
\textit{permettre}. De la même manière, en sélectionnant le mot \textit{justifier}, les liens reliant
l'arbitre Solensean aux autres arbitres sont supprimés. L'arbitre Solensean est le seul
à ne pas utiliser le terme \textit{justifier}.

En utilisant ce filtrage sur tous les termes du lexique des arbitres, nous constatons
que les arbitres qui ont participé à au moins 50 votes d'arbitrage, possèdent un très
large vocabulaire en commun. Les différences que nous avons notées concernent
l'arbitre R qui n'utilise pas les termes \textit{attendre, permettre, contributeurs, fond,
prendre}, et \textit{déjà}, contrairement aux autres arbitres, et l'arbitre Solensean qui est le
seul à ne pas utiliser le mot \textit{justifier}.

\subsection{Visualisation des propriétés d'un nœud}

\begin{figure*}[!htb]
 \centering
 \includegraphics[width=.8\textwidth]{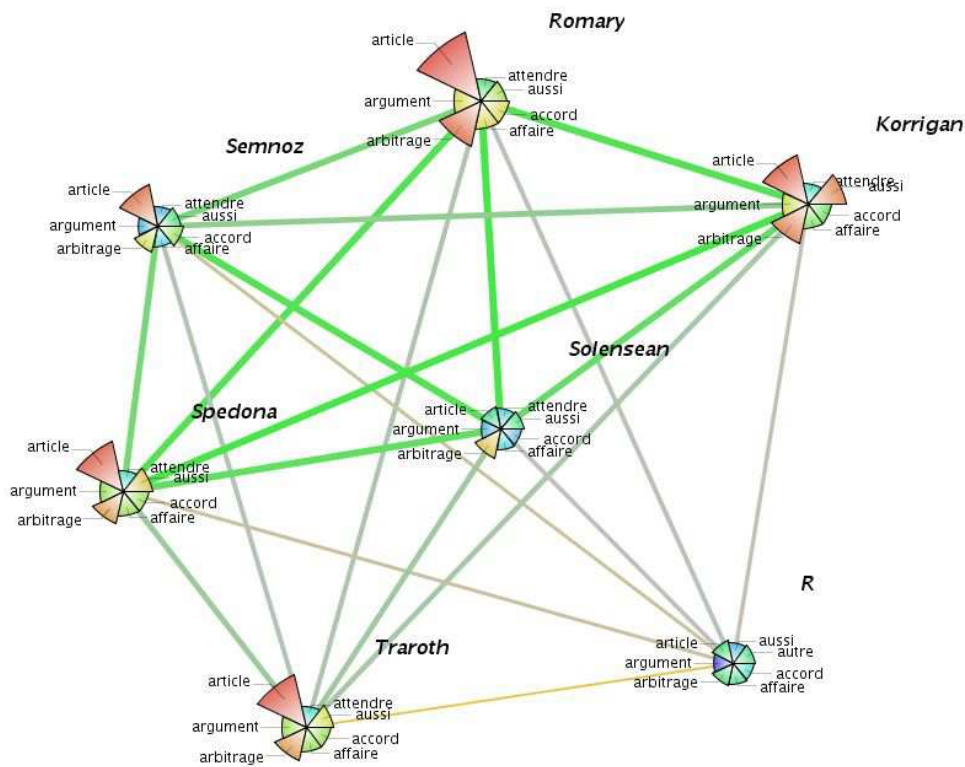}
 \caption{Distribution des 7 mêmes termes pour chaque arbitre ayant participé à au
moins 50 votes.}
 \label{fig:5-arbitresTermes}
\end{figure*}

On peut également visualiser, pour chaque arbitre, la distribution du vocabulaire
qu'il utilise. La Figure \ref{fig:5-arbitresTermes} montre la distribution d'un même ensemble de 7 termes
(sauf pour l'arbitre R qui n'utilise pas l'un des 7 mots, \textit{attendre}) pour chaque arbitre.
On constate que les arbitres R et Solensean ont un profil de lexique différent de celui
des autres.

\section{Visualisation croisée entre deux réseaux}

\begin{figure*}[!htb]
 \centering
 \includegraphics[width=.8\textwidth]{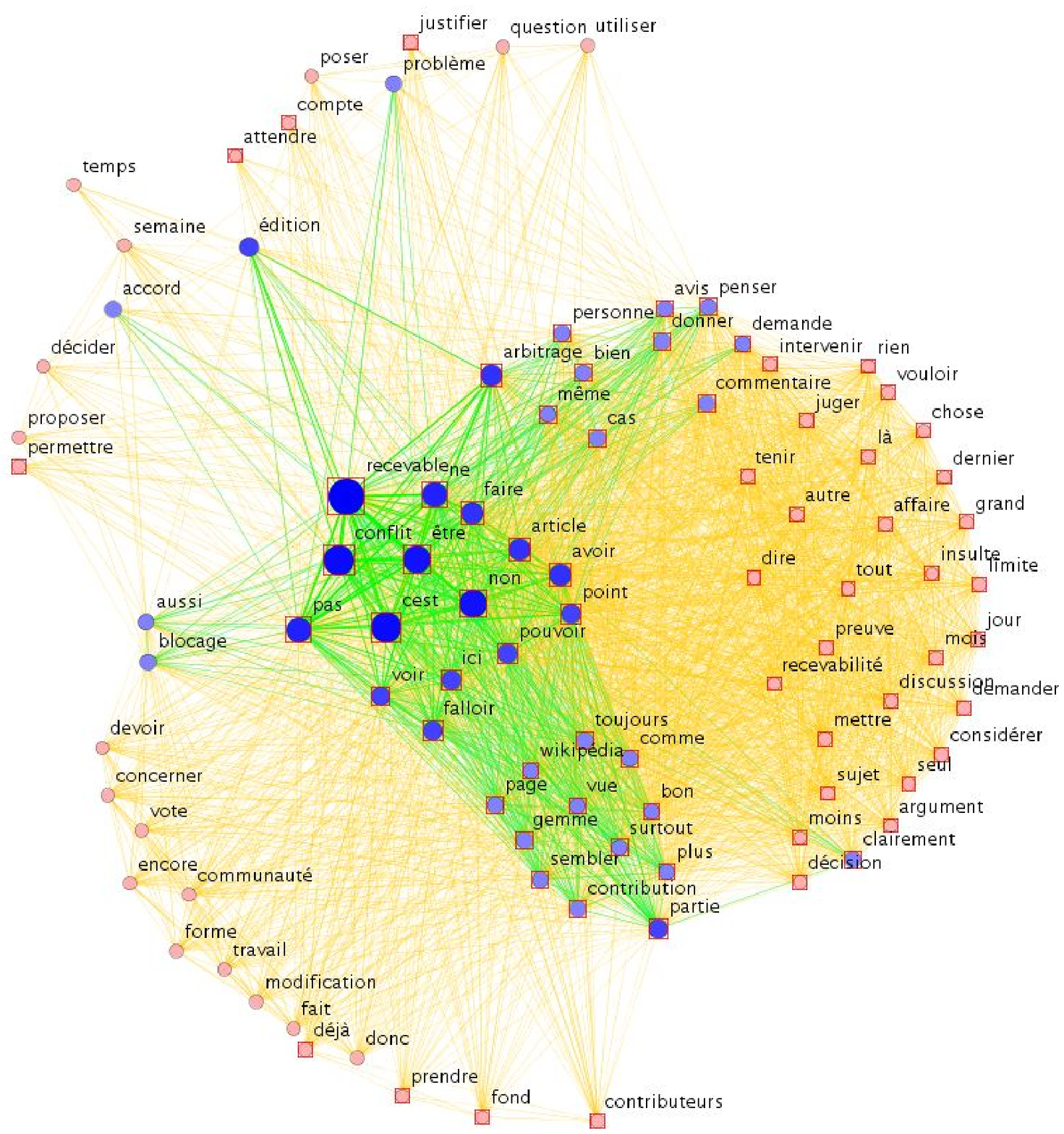}
 \caption{Les nœuds-termes utilisés par l'arbitre Aoineko (entourés d'un carré), 
 dans le lexique des arbitres.}
 \label{fig:6-aoinekoTermes}
\end{figure*}

La sélection d'un nœud-terme du réseau lexical met visuellement en évidence
dans le réseau social tous les nœuds-arbitre qui utilisent ce mot. Inversement, la
sélection d'un nœud-arbitre du réseau social met en évidence dans le réseau lexical
tous les nœuds-terme utilisés par cet arbitre. La sélection initiale d'un nœud dans l'un
des réseaux est marquée par une couleur particulière sur le nœud, les mises en
évidence en réaction dans l'autre réseau sont marqués par un carré sur le nœud\footnote{Il est 
possible d'effectuer ses propres choix de couleurs par un paramétrage personnalisé dans l'interface.}. Par
exemple, la Figure \ref{fig:6-aoinekoTermes} montre les termes utilisés par l'arbitre Aoineko, sélectionné dans
le réseau social.

Cet arbitre utilise un large vocabulaire: on voit en effet que très peu de nœuds-
termes ne sont pas entourés d'un carré. En revanche, la Figure \ref{fig:2-arbitresVoc}, qui met en évidence
les nœuds-termes utilisés par l'arbitre Greudin (entourés d'un carré), montre que
ceux-ci sont moins nombreux. On peut ainsi comparer visuellement les tailles des
vocabulaires utilisés par les différents arbitres.

\section{Conclusion}

Le logiciel GraphDuplex, par un ensemble de paramétrages de visualisation et de
filtrage, associés à des méthodes statistiques et des méthodes de calcul sur des
graphes, permet une exploration interactive de plusieurs réseaux. Ce logiciel,
développé en Java, utilise deux librairies opensource Jung \citep{OMadadhainAl2003} 
et InfoVis Toolkit \citep{Fekete2004}, ainsi que des composants Semiophore. Il
exploite également Graphviz d’ATT\footnote{\href{http://www.graphviz.org}{\url{http://www.graphviz.org}}}. La mise en correspondance des attributs typés
dans une base de données et des attributs graphiques dans les graphes ont été
spécifiquement développés pour GraphDuplex. Cela passe par une IHM qui donne la
possibilité de régler les paramètres graphiques en fonction des attributs (champs de
la base de données). La facilité à se brancher sur n'importe quelle base provient d'une
couche abstraction qui décrit dans un modèle XML les liens entre les données et la
morphologie des graphes (nœuds, liens) ainsi que les liens entre les différents
graphes (contraintes de sélections transversales).

GraphDuplex est particulièrement intéressant si l'on veut analyser un réseau
lexical lié à une activité sociale, mais il peut être utilisé à d'autres fins, comme par
exemple la comparaison de deux réseaux lexicaux sur un même thème, à des temps
différents. Pour montrer les possibilités du logiciel nous avons pris l'exemple du
réseau social des arbitres au Comité d'arbitrage de Wikipédia, associé au réseau
lexical de leurs interventions dans ce même comité. En ce qui concerne le réseau
lexical, nous avons pu mettre en évidence les différences de vocabulaire entre les
arbitres (Figures \ref{fig:4-arbitresNetPermettre} et \ref{fig:5-arbitresTermes}), les différences entre les distributions d'un même ensemble
de mots pour différents individus du réseau social, ou bien encore les thèmes les plus
fréquents. Les requêtes dynamiques inter-réseaux permettent aussi de repérer les
individus du réseau social qui utilisent les termes et les thèmes mis en évidence dans
le réseau lexical.

Pour compléter cette étude, un autre réseau social pourrait être ajouté, celui des
contributeurs à Wikipédia qui comparaissent devant le Comité d'arbitrage. Trois
réseaux liés entre eux seraient ainsi visualisés simultanément: le réseau social des
arbitres, le réseau social des protagonistes des conflits, et le réseau lexical des
interventions de l'ensemble des individus des deux réseaux sociaux. Des
comparaisons pourraient ainsi être faites entre les vocabulaires des arbitres et des
contributeurs protagonistes des conflits.

\section*{Remerciements}

Ce travail a été réalisé dans le cadre du projet Autograph ANR-05-RNRT-03002
(S0604108 W)

\bibliographystyle{apalike-fr}
\bibliography{\mabiblio}

\end{document}